\documentclass[epj,nopacs]{svjour}

\usepackage{amssymb,amsfonts,amsmath,citesort}

\usepackage{graphicx}
\usepackage{dcolumn}
\usepackage{amsmath}
\usepackage[notheorems]{definitions}
\usepackage{xspace}
\usepackage[dvips]{color}
\usepackage{rotating}
\usepackage{multirow}
\usepackage{algorithm,algorithmic}

\begin{document}

\setlength{\parskip}{0pt}
\setlength{\tabcolsep}{2pt}
\setlength{\arraycolsep}{2pt}

\newcommand{\dkedit}[1]{{\color{blue}\hspace{0ex} #1}}
\newcommand{\gaedit}[1]{{\color{green}\hspace{0ex} #1}}
\newcommand{\prune}{}
\newcommand{\remove}[1]{{\color{red}\hspace{0ex} #1}}



\newcommand{\MAX}{\ensuremath{\mathrm{MAX}}\xspace}
\newcommand{\Deg}[1]{\ensuremath{d_{#1}}\xspace}
\newcommand{\MAXAGREE}{{\sc MaxAgree}\xspace}
\newcommand{\MINDISAGREE}{{\sc MinDisagree}\xspace}
\newcommand{\MOD}{\ensuremath{Q}\xspace}
\newcommand{\Mod}[1]{\ensuremath{\MOD(#1)}\xspace}
\newcommand{\ModChange}[1]{\ensuremath{\Delta \MOD(#1)}\xspace}
\newcommand{\MODMAT}{\ensuremath{M}\xspace}
\newcommand{\AdjM}[2]{\ensuremath{a_{#1,#2}}\xspace}
\newcommand{\ModM}[2]{\ensuremath{m_{#1,#2}}\xspace}
\newcommand{\KRONECKER}{\ensuremath{\delta}\xspace}
\newcommand{\Kronecker}[2]{\ensuremath{\KRONECKER(#1,#2)}\xspace}
\newcommand{\NClust}[1]{\ensuremath{\gamma(#1)}\xspace}
\newcommand{\CLUSTERING}{\ensuremath{\mathcal{C}}\xspace}
\newcommand{\Clust}[1][]{\ensuremath{\ifthenelse{\equal{#1}{}}{C}{C_{#1}}}\xspace}
\newcommand{\ClustP}[1][]{\ensuremath{\ifthenelse{\equal{#1}{}}{C'}{C'_{#1}}}\xspace}
\newcommand{\ClustPP}[1][]{\ensuremath{\ifthenelse{\equal{#1}{}}{C''}{C''_{#1}}}\xspace}
\newcommand{\LPLUS}{`$+$'\xspace}
\newcommand{\LMINUS}{`$-$'\xspace}


\title{Modularity-Maximizing Graph Communities via Mathematical Programming}
\author{Gaurav Agarwal\inst{1} \and David Kempe\inst{2}}
\institute{Google Inc. \and Computer Science Department, University of Southern California, LosAngeles, CA90089}


\abstract{
In many networks, it is of great interest to identify \todef{communities}, 
unusually densely knit groups of individuals. Such communities often shed light 
on the function of the networks or underlying properties of the individuals. 
Recently, Newman suggested \todef{modularity} as a natural measure of the 
quality of a network partitioning into communities. 
Since then, various algorithms have been proposed for (approximately)
maximizing the modularity of the partitioning determined.
In this paper, we introduce the technique of rounding mathematical
programs to the problem of modularity maximization, presenting two
novel algorithms. 
More specifically, the algorithms round solutions to linear and vector
programs. Importantly, the linear programing algorithm comes with an a
posteriori approximation guarantee: by comparing the solution quality
to the fractional solution of the linear program, a bound on the
available ``room for improvement'' can be obtained.
The vector programming algorithm provides a similar bound for the best
partition into \emph{two} communities.
We evaluate both algorithms using experiments on several standard test
cases for network partitioning algorithms, and find that they perform
comparably or better than past algorithms.
}

\maketitle

\section {INTRODUCTION}
Many naturally occurring systems of interacting entities can be conveniently 
described using the notion of networks. \todef{Networks} (or \todef{graphs}) 
consist of \todef{nodes} (or \todef{vertices}) and \todef{edges} between them 
\cite{newman:barabasi:watts}. For example, \todef{social networks} 
\cite{scott:social-network,wasserman:faust} describe individuals and their 
interactions, such as friendships, work relationships, sexual contacts, etc. 
Hyperlinked text, such as the World Wide Web, consists of pages and their 
linking patterns \cite{kleinberg:kumar:raghavan:rajagopalan:tomkins}. Metabolic 
networks model enzymes and metabolites with their reactions 
\cite{guimera:amaral}.

In analyzing and understanding such networks, it is frequently extremely useful 
to identify \todef{communities}, which are informally defined as ``unusually 
densely connected sets of nodes''. Among the benefits of identifying community 
structure are the following:

\begin{enumerate}
\item Frequently, the nodes in a densely knit community share a salient 
real-world property. For social networks, this could be a common 
interest or location;
for web pages, a common topic or language; and for 
biological networks, a common function. Thus, by analyzing structural features 
of a network, one can infer semantic attributes.

\item By identifying communities, one can study the communities 
individually. Different communities often exhibit significantly 
different properties, making a global analysis of the network inappropriate. 
Instead, a more detailed analysis of individual communities leads to more 
meaningful insights, for instance into the roles of individuals.

\item Conversely, each community can be compressed into a single ``meta-node'', 
permitting an analysis of the network at a coarser level, and 
a focus on higher-level structure. This approach can also be useful in 
visualizing an otherwise too large or complex network.
\end{enumerate}

For a much more detailed discussion of these and other motivations,
see for instance \cite{newman:eigenvectors}. 
Due to the great importance of identifying 
community structure in graphs, there has been a large amount of work in 
computer science, physics, economics, and sociology (for some examples, see 
\cite{newman:eigenvectors,flake:lawrence:giles:coetzee,girvan:newman,newman:detecting-structure,duch:arenas}). 
At a very high level, one can identify two lines of work. In one line 
\cite{flake:lawrence:giles:coetzee,flake:tarjan:tsioutsiouliklis}, dense 
communities are identified one at a time, which allows vertices to be part of 
multiple communities. Depending on the context, this may or may not be 
desirable.
Often, the communities identified will correspond to 
some notion of ``dense subgraphs'' 
\cite{flake:lawrence:giles:coetzee,flake:tarjan:tsioutsiouliklis,UnbalancedCuts,charikar:greedy-dense}.

An alternative is to seek a \emph{partition} of the graph into disjoint 
communities, i.e., into sets such that each node belongs to exactly one set. 
This approach is preferable when a ``global view'' of the network is 
desired, and is the one discussed in the present work. It is closely related to 
the problem of \todef{clustering}; indeed, ``graph clustering'',
``partitioning'',  and ``community identification'' are
often, including here, used interchangeably. 

Many approaches have been proposed for finding such partitions, 
based on spectral properties, flows, edge agglomeration, and many 
others (for a detailed overview and comparison, see
\cite{newman:eigenvectors}). 
The approaches differ in whether or not 
a hierarchical partition (recursively subdividing communities
into sub-communities) is sought, whether the number of
communities or their size is pre-specified by the user or decided by
the algorithm, as well as other parameters. For a survey, see
\cite{newman:detecting-structure}.

A particularly natural approach was recently proposed by Newman and Girvan 
\cite{newman:girvan:finding-evaluating,newman:fast-community}. Newman 
\cite{newman:fast-community} proposes to find a community partition maximizing 
a measure termed \todef{modularity}. The modularity of a given
clustering is the number 
of edges inside clusters (as opposed to crossing between clusters), minus the 
expected number of such edges if the graph were random conditioned on its 
degree distribution \cite{newman:girvan:finding-evaluating} .
Subsequent work by 
Newman et al.~and others has shown empirically that modularity-maximizing 
clusterings often identify interesting community structure in real networks, 
and focused on different heuristics for obtaining such clustering 
\cite{newman:eigenvectors,duch:arenas,newman:girvan:finding-evaluating,newman:fast-community,clauset:newman:moore,clauset:local-community,newman:modularity-community}.
For a detailed overview and comparison of many of the proposed
heuristics for modularity maximization, see \cite{danon:duch:diaz-guilera:arenas}.

\begin{remark}
It should be noted that
graph communities found by maximizing modularity should be judged carefully. 
While modularity is one natural measure of community structure in networks, 
there is no guarantee that it captures the particular structure relevant in a 
specific domain. For example, Fortunato and Barth\'{e}lemy 
\cite{fortunato:barthelemy} have recently shown that modularity and more 
generally, each
``quality function'' (characterizing the quality of the entire partition in one 
number) have an intrinsic resolution scale, and can therefore fail to detect 
communities smaller than that scale. More fundamentally, Kleinberg 
\cite{kleinberg:clustering-impossibility} has shown that no single clustering 
method can ever satisfy four natural desiderata on all instances.
\end{remark}

Recently, Brandes et al.~\cite{BDGGHNW:journal} have shown that finding the clustering 
of maximum modularity for a given graph
is NP-complete. This means that efficient algorithms to always find an
optimal clustering, in time  polynomial in the size of the graph for
\emph{all} graphs, are unlikely to exist. 
It is thus desirable to develop heuristics yielding 
clusterings as close to optimal as possible. 

In this paper, we introduce the technique of solving and rounding
fractional mathematical programs to the problem of community
discovery, and propose two new algorithms for finding
modularity-maximizing clusterings.
The first algorithm is based on a linear programming (LP) relaxation of 
an integer programming (IP) formulation.
The LP relaxation will put nodes ``partially in the same cluster''. 
We use a ``rounding'' procedure due to Charikar et
al.~\cite{charikar:guruswami:wirth} for the problem of
\todef{Correlation Clustering}
\cite{bansal:blum:chawla:correlation-clustering}. The idea of the 
algorithm is to interpret ``partial membership of the same cluster'' as a 
distance metric, and group together nearby nodes.

The second algorithm is based on a vector programming (VP) relaxation of a 
quadratic program (QP). It recursively splits one partition into two smaller 
partitions while a better modularity can be obtained. 
It is similar in spirit to an approach recently 
proposed by Newman \cite{newman:eigenvectors,newman:modularity-community}, 
which repeatedly divides clusters based on the first eigenvector of the 
modularity matrix.
Newman's approach can be thought of as embedding nodes in 
the interval $[-1,1]$, and then cutting the interval in the middle.
The VP embeds nodes on the surface of a high-dimensional hypersphere,
which is then randomly cut into two halves containing the
nodes. The approach is thus very similar to the algorithm for Maximum
Cut due to Goemans and Williamson \cite{goemans:williamson:max-cut}.

A significant advantage of our algorithms over past approaches is that
they come with an \emph{a posteriori} error bound. The value obtained
by the LP relaxation is an upper bound on the maximum achievable
modularity. Although in principle, this bound could be
loose, it was very accurate in all our test instances. 
By comparing the modularity obtained by an algorithm 
against the LP value, we can estimate how 
close to optimal the solution is. Similarly, the value of the VP relaxation 
gives a bound on the best division of the graph into \emph{two} communities.

We evaluate our algorithms on several standard test cases for graph community 
identification.   
On every test case where an upper bound on the optimal solution could be 
determined, the solution found using both our algorithms attains at 
least 99\% of the theoretical upper bound; sometimes, it is optimal. 
In addition, both algorithms match or outperform past modularity
maximization algorithms on most test cases.
Thus, our results suggest that these algorithms are excellent choices
for finding graph communities.

The performance of our algorithms comes at a price of 
significantly slower running time and higher memory requirements. The bulk of 
both time and memory are consumed by the LP or VP solver; the rounding is 
comparatively simple. Mostly due to the high memory requirements, the LP 
rounding algorithm can currently only be used on networks of up to a few 
hundred nodes. The VP rounding algorithm has lower running time and memory 
requirements than the LP method and scales to networks of up to a few 
thousand nodes on a personal desktop computer.

We believe that despite their lower efficiency,
our algorithms provide three important contributions. First, they are
the first algorithms with guaranteed polynomial running time
to provide a posteriori performance guarantees. 
Second, they match or outperform past
algorithms for medium-sized networks of practical interest. 
And third, the approach proposed in our paper 
introduces a new algorithmic paradigm to the physics
  community. Future work using these techniques would have the
  potential to produce more
efficient algorithms with smaller resource requirements. 
Indeed, in the past, algorithms
based on rounding LPs were often a first step towards achieving the
same guarantees with purely combinatorial algorithms.
Devising such algorithms is a direction of ongoing work.

\section{Preliminaries} \label{sec:preliminaries}
The network is given as an undirected graph $G=(V,E)$. The adjacency matrix of 
$G$ is denoted by $A = (\AdjM{u}{v})$: thus, $\AdjM{u}{v} = \AdjM{v}{u} = 1$ if $u$ and $v$ 
share an edge, and $\AdjM{u}{v} = \AdjM{v}{u} = 0$ otherwise. The degree of a node $v$ is 
denoted by \Deg{v}. A \todef{clustering} $\CLUSTERING = \SET{\Clust[1], \ldots, 
\Clust[k]}$ is a partition of $V$ into disjoint sets \Clust[i]. We use 
\NClust{v} to denote the (unique) index of the cluster that node $v$ belongs to.

The \todef{modularity} \cite{newman:girvan:finding-evaluating} of a clustering 
\CLUSTERING is the total number of edges inside clusters, minus the expected 
number of such edges if the graph were a uniformly random
  multigraph subject to its degree sequence. In order to be able to
compare the modularity for graphs of different  
sizes, it is convenient to normalize this difference by a factor of $1/2m$, so 
that the modularity is a number from the interval $[-1,1]$.

If nodes $u,v$ have degrees $\Deg{u}, \Deg{v}$, then any one of the $m$ edges 
has probability $2 \frac{\Deg{u}}{2m} \cdot \frac{\Deg{v}}{2m}$ of connecting 
$u$ and $v$ (the factor 2 arises because either endpoint of the edge could be 
$u$ or $v$). By linearity of expectation, the expected number of edges between 
$u$ and $v$ is then $\frac{\Deg{u}\Deg{v}}{2m}$. Thus, the modularity of a 
clustering \CLUSTERING is
\begin{eqnarray}
\Mod{\CLUSTERING} 
& := & \frac{1}{2m} \sum_{u,v} (\AdjM{u}{v} - \frac{\Deg{u}\Deg{v}}{2m}) \cdot
\Kronecker{\NClust{u}}{\NClust{v}}, \label{eqn:modularity}
\end{eqnarray}
where \KRONECKER denotes the Kronecker Delta,
which is 1 iff its arguments are identical, and 0 otherwise.
Newman \cite{newman:eigenvectors} terms the matrix \MODMAT with entries
$\ModM{u}{v} := \AdjM{u}{v} - \frac{\Deg{u}\Deg{v}}{2m}$ the 
\todef{modularity matrix} of $G$. For a more detailed discussion of
the probabilistic interpretation of modularity and generalizations of
the measure, see the recent paper by Gaertler et
al.~\cite{gaertler:goerke:wagner:significance}.

\section{Algorithms} \label{sec:algorithms}
\subsection{Linear Programming based algorithm}
\subsubsection{The Linear Program}
Based on Equation \ref{eqn:modularity}, we can phrase the modularity 
maximization problem as an integer linear program (IP). (For an introduction to 
Linear Programming, we refer the reader to
\cite{chvatal:linear-programming,karloff:linear-programming};
for the technique of LP rounding, see 
\cite{vazirani:approximation-algorithms}.) The linear program has one 
variable $x_{u,v}$ for each pair $(u,v)$ of vertices. We interpret 
$x_{u,v} = 0$ to mean that $u$ and $v$ belong to the same cluster, and $x_{u,v} 
= 1$ that $u$ and $v$ are in different clusters. Then, the objective function 
to be maximized can be written as $\sum_{u,v} \ModM{u}{v} (1-x_{u,v})$. 
This is a linear function, because the \ModM{u}{v} are constants. 
We need to ensure that 
\prune{the $x_{u,v}$ are consistent with each other:}
if $u$ and $v$ are in the same cluster, and $v$ and $w$ 
are in the same cluster, then so are $u$ and $w$. This constraint can be 
written as a linear inequality $x_{u,w} \leq x_{u,v} + x_{v,w}$. It is not 
difficult to see that the $x_{u,v}$ are consistent (i.e., define a clustering) 
if and only if this inequality holds for all triples $(u,v,w)$. Thus, we obtain 
the following integer linear program (IP):

\begin{equation}
\begin{array}{lll} 
\mbox{Maximize} & \multicolumn{2}{l}{\frac{1}{2m} \cdot \sum_{u,v}{\ModM{u}{v} \cdot (1-x_{u,v})}}\\
\mbox{subject to} & x_{u,w} \leq x_{u,v} + x_{v,w} & \mbox{ for all } u,v,w\\
& x_{u,v} \in \SET{0,1} &  \mbox{ for all } u,v
\end{array}
\label{eqn:LP}
\end{equation}

Solving IPs is also NP-hard,
and thus unlikely to be possible in polynomial time. However, by 
replacing the last constraint 
--- that each $x_{u,v}$ be an integer from \SET{0,1} --- 
with the constraint that each $x_{u,v}$ be a real 
number between 0 and 1, we obtain a linear program (LP). LPs can be 
solved in polynomial time
\cite{karloff:linear-programming,karmarkar:linear-programming}, and
even quite efficiently in practice. (For our experiments, we use the
widely used commercial package CPLEX.) The downside is that the
solution, being fractional, does not correspond to a clustering. As a
result, we have to apply a post-processing step, called ``rounding'' of the LP.

\subsubsection{The LP Rounding Algorithm}
Our LP rounding algorithm is essentially identical to one proposed by Charikar
et al.~\cite{charikar:guruswami:wirth} for the \todef{Correlation 
Clustering} problem \cite{bansal:blum:chawla:correlation-clustering}. In 
correlation clustering, one is given an undirected graph $G=(V,E)$
with each edge labeled either \LPLUS (modeling similarity between endpoints) or \LMINUS 
(modeling dissimilarity). The goal is to partition the graph into clusters such 
that few vertex pairs are classified incorrectly. Formally, 
in the \MINDISAGREE version of the problem,
the goal is to minimize
the number of \LMINUS edges inside clusters plus the
number of \LPLUS edges between clusters.
In the \MAXAGREE version, which is not as relevant to our
approach, the goal is to maximize the number of \LPLUS edges inside
clusters plus the number of \LMINUS edges between clusters.
Using the same 0-1 variables $x_{u,v}$ as we did above, Charikar et 
al.~\cite{charikar:guruswami:wirth} formulate \MINDISAGREE as 
follows:

\begin{displaymath}
\begin{array}{lll}
\mbox{Minimize} & \multicolumn{2}{l}{\sum_{(u,v) \in E_+} x_{u,v} + \sum_{(u,v) \in E_-} (1 - x_{u,v})} \\
\mbox{subject to} & x_{u,w} \leq x_{u,v}+x_{v,w} & \mbox{ for all } u,v,w\\
& x_{u,v} \in \SET{0,1} & \mbox{ for all } u,v,
\end{array}
\end{displaymath}
where $E_+$ and $E_-$ denote the sets of edges labeled \LPLUS and \LMINUS, 
respectively.
The objective can be rewritten as 
$\SetCard{E_+} - \sum_{(u,v) \in E} \mu_{u,v} (1-x_{u,v})$, 
where $\mu_{u,v}$ is 1 for \LPLUS edges and -1 for \LMINUS edges. 
The objective is minimized when 
$\sum_{(u,v) \in E} \mu_{u,v} (1-x_{u,v})$ is maximized;
thus, except for the shift by the constant \SetCard{E_+}, 
\MINDISAGREE takes on the same form as 
modularity maximization with $\ModM{u}{v} = \mu_{u,v}$.

The rounding algorithm proposed by Charikar et
al.~\cite{charikar:guruswami:wirth} comes with an \emph{a priori}
error guarantee that the objective produced is never more than 4 times
the optimum. Algorithm with such guarantees are called
\emph{Approximation Algorithms}
\cite{vazirani:approximation-algorithms}, and it would be desirable to
design such algorithms for Modularity Maximization as well. 
Unfortunately, the shift by a constant prevents the approximation
guarantees from \cite{charikar:guruswami:wirth} from carrying over to
the Modularity Maximization problem. However, the analogy suggests that
algorithms for rounding the  solution to the \MINDISAGREE LP may
perform well in practice for Modularity Maximization.

Our rounding algorithm, based on the one by Charikar et al., first solves the 
 linear program (\ref{eqn:LP}) without the integrality constraints. 
This leads to a \emph{fractional} assignment $x_{u,v}$ for every pair of 
vertices. The LP constraints, applied to fractional values $x_{u,v}$,
exactly correspond to the triangle inequality. Hence, the $x_{u,v}$ form a 
metric, and we can interpret them as ``distances'' between the vertices. We use 
these distances to repeatedly find clusters of ``nearby'' nodes, which are then 
removed. The full algorithm is as follows:

\begin{algorithm}[H]
\caption{Modularity Maximization Rounding}
\begin{algorithmic}[1] \STATE Let $S = V$.
	\WHILE{$S$ is not empty}
		\STATE Select a vertex $u$ from $S$. 
		\STATE Let $T_u$ be the set of vertices whose distance
                from $u$ is at most \half.
		\IF{the average distance of the vertices in $T_u \setminus \SET{u}$ from $u$ is less than \quarter}
			\STATE Make $C = T_u$ a cluster. 
		\ELSE 
			\STATE Make $C = \SET{u}$ a singleton cluster. 
		\ENDIF 
		\STATE Let $S = S \setminus C$. 
	\ENDWHILE
\end{algorithmic}
\end{algorithm}

Step 3 of the rounding algorithm is underspecified: it does not say 
\emph{which} of the remaining vertices $u$ to choose as a center
next. We found that selecting a random center in each iteration, and
keeping the best among $1000$ independent executions of the entire
rounding algorithm, significantly outperformed two natural
alternatives, namely selecting the largest or smallest cluster. 
\prune{In particular, selecting the largest cluster is a 
significantly inferior heuristic.}

%

As a post-processing step to the LP rounding, we run a
\emph{local-search} algorithm proposed by Newman
\cite{newman:eigenvectors} to refine the results further.
The post-processing step is briefly described below.


An important benefit of the LP rounding 
method is that it provides an upper bound on the best solution. For
the best clustering is the optimum solution to the integer LP 
(\ref{eqn:LP}); removing the integrality constraint can only increase the set 
of allowable solutions to the LP, improving the objective value 
that can be obtained. The upper bound enables us to lower-bound the performance 
of clustering algorithms.

The other useful feature of our algorithm is its inherent capability to find 
different clusterings with similar modularity. The randomization naturally 
leads to different solutions, of which several with highest modularity values 
can be retained, to provide a more complete picture of possible
cluster boundaries.



\subsection{Vector Program Based Algorithm}
In this section, we present a second algorithm which is more efficient
in practice, at the cost of slightly reduced performance. 
It produces a ``hierarchical clustering'', in the sense that the
clustering is obtained by repeatedly finding a near-optimal division
of a larger cluster. For two reasons, this clustering is not truly
hierarchical: 
First, we do not seek to optimize a global function of the
entire hierarchy, but rather optimize each split locally. Second, we
again apply a local search based post-processing step to improve the
solution, thus rearranging the clusters.
 Despite multiple recently proposed hierarchical clustering
algorithms (e.g.,
\cite{girvan:newman,newman:eigenvectors,sales-pardo:guimera:moreira:amaral}),
there is far from general agreement on what objective functions would
capture a ``good'' hierarchical clustering. Indeed, different
objective functions can lead to significantly different clusterings.
While our clustering is not truly hierarchical, the order and position
of the splits that it produces still reveal much high-level
information about the network and its clusters.

As discussed above, our approach is to
aim for the best division \emph{at each level individually}, 
requiring a partition into two clusters at each level. 
Clusters are recursively subdivided as long as an
improvement is possible. Thus, a solution hinges on being able to find
a good partition of a given graph into \emph{two} communities.  
The LP rounding algorithm presented in the previous
section is not applicable to this problem, as it does not permit
specifying the number of communities. Instead, we will use a Vector
Programming (VP) relaxation of a Quadratic Program (QP) to find a good
partition of a graph into two communities. 

\subsubsection{The Quadratic Program}
Our approach is motivated by the same observation that led Newman 
\cite{newman:eigenvectors} to an eigenvector-based partitioning approach.
For every vertex $v$, we have a variable $y_v$ which is 1 or -1 
depending on whether the vertex is in one or the other partition. Since each
pair $u,v$ adds \ModM{u}{v} to the objective iff $u$ and $v$ are in
the same partition (and zero otherwise), the objective function can be 
written as $ \frac{1}{4m} \sum_{u,v} \ModM{u}{v} (1+y_u y_v)$. Newman 
\cite{newman:eigenvectors} rewrites this term further as $\frac{1}{4m} 
\Transpose{\mathbf{y}} \MODMAT \mathbf{y}$ (where $\mathbf{y}$ is the vector of 
all $y_v$ values), and observes that if the entries $y_v$ were not restricted 
to be $\pm 1$, then the optimal $\mathbf{y}$ would be the principal eigenvector 
of \MODMAT. His approach,
in line with standard spectral partitioning approaches (e.g.,
\cite{fiedler:eigenvectors}), is then to compute the 
principal eigenvector $\mathbf{y}$, and partition the nodes into 
positive  $y_v$ and negative $y_v$. Thus, in a
sense, Newman's approach can be considered as embedding the nodes
optimally on the line, and then rounding the fractional
solution into nodes with positive and negative coordinates. 

Our solution also first embeds the nodes into a metric space, and then rounds 
the locations to obtain two communities. However, it is motivated by 
considering the objective function as a strict quadratic program (see, e.g., 
\cite{vazirani:approximation-algorithms}). We can write the 
problem of partitioning the graph into two communities of maximum modularity as

\begin{equation}
 \begin{array}{lll}
  \mbox{Maximize} & \multicolumn{2}{l}{\frac{1}{4m} \sum_{u,v} \ModM{u}{v} \cdot (1 + y_u  y_v)}\\
  \mbox{subject to} & y_v^2 = 1 & \mbox{ for all } v.
 \end{array}
\label{eqn:QP}
\end{equation}
  Notice that the constraint $y_v^2 = 1$ ensures that each $y_v$ is $\pm 1$ in 
  a solution to (\ref{eqn:QP}).

Quadratic Programming, too, is NP-complete.
Hence, we use the standard technique of relaxing the QP 
(\ref{eqn:QP}) to a corresponding Vector Program (VP), which in turn can be 
solved in polynomial time using semi-definite programming (SDP). To turn a 
strict quadratic program into a vector program, one replaces each variable 
$y_v$ with a ($n$-dimensional) vector-valued variable $\mathbf{y}_v$, and each 
product $y_u y_v$ with the inner product $\mathbf{y}_u \cdot \mathbf{y}_v$.
We use the standard process \cite{vazirani:approximation-algorithms}
for transforming the VP formulation to the SDP formulation and for
obtaining back the solution to the VP from the solution to SDP. 
For solving the SDP problems in our experiments, we use a standard 
off-the-shelf solver CSDP \cite{borchers:csdp}.

The result of solving the VP will be vectors $\mathbf{y}_v$ for all vertices 
$v$, which can be interpreted as an embedding of the nodes on the surface of 
the hypersphere in $n$ dimensions. (The constraint $\mathbf{y}_v \cdot 
\mathbf{y}_v = 1$ for all $v$ ensures that all nodes are embedded at distance 1 
from the origin.) Thus, the inner product of two node positions $\mathbf{y}_u, 
\mathbf{y}_v$ is equal to the cosine of the angle between them. As a result, 
the optimal VP solution will ``tend to'' have node pairs with negative 
\ModM{u}{v} far apart (large angles), and node pairs with positive \ModM{u}{v} 
close (small angles).

\subsubsection{Rounding the Quadratic Program}
To obtain a partition from the node locations $\mathbf{y}_v$, we use a 
rounding procedure proposed by Goemans and Williamson 
\cite{goemans:williamson:max-cut} for the Max-Cut problem. In the Max-Cut 
problem, an undirected graph is to be partitioned into two disjoint node 
sets so as to maximize the number of edges crossing between them. 
This objective can be written as a quadratic program as follows (notice the 
similarity to the Modularity Maximization QP):

\[  \begin{array}{lll} \mbox{Maximize} & \multicolumn{2}{l}{\frac{1}{2} \sum_{(u,v) \in E} (1 - y_u  y_v)}\\
\mbox{subject to} & y_v^2 = 1 & \mbox{ for all } v. \end{array} \]

The rounding procedure of Goemans and Williamson 
\cite{goemans:williamson:max-cut}, which we adopt here, chooses a random 
$(n-1)$-dimensional hyperplane passing through the origin, and uses the 
hyperplane to cut the hypersphere into two halves. The two partitions are 
formed by picking the vertices lying on each side of the hypersphere. The 
cutting hyperplane is represented by its normal vector $\mathbf{s}$, 
which is an $n$-dimensional vector, each of whose components is an 
independent $\mathcal{N}(0,1)$ Gaussian. (It is well known and easy to verify 
that this makes the direction of the normal uniformly random.) To cut the 
hypersphere, we simply define $S := \Set{v}{\mathbf{y}_v \cdot \mathbf{s} \geq 
0}$ and $\Compl{S} := \Set{v}{\mathbf{y}_v \cdot \mathbf{s} < 0}$. 
Once the VP has been solved (which is the expensive part), one can
easily choose multiple random hyperplanes, and retain the best
resulting partition. In our experiments, we chose the best of 5000
hyperplanes.

A different approach to rounding VP solutions of the form
(\ref{eqn:QP}) was recently proposed by Charikar and Wirth
\cite{charikar:wirth}, again in the context of Correlation
Clustering. Their method first projects the hypersphere on a random
line, scales down large coordinates, and then rounds randomly.
Their method gives an a priori error guarantee of $\Omega(1/\log n)$
under the assumption
that all diagonal entries of the matrix $M$ are zero. In fact, if the
matrix is also positive semi-definite, then a result of Nesterov
\cite{nesterov:semidefinite} shows that the approximation guarantee
can be improved to $2/\pi$. Unfortunately, the modularity matrix $M$
is neither positive semi-definite nor does it have an all-zero trace;
hence, neither of these approximation results is applicable to the
problem of finding the modularity-maximizing partition into two
communities.

We also implemented the rounding procedure of \cite{charikar:wirth}, and tested
it on the same example networks as the other algorithms. We found that its
performance is always inferior to the hyperplane based algorithm, sometimes
significantly so. Since the algorithm is not more efficient, either, we omit
the results from our comparison in Section \ref{sec:experiments}.

\subsubsection{The Hierarchical Clustering Algorithm}
Note that the effect of partitioning a community \Clust further 
into two sub-communities $\ClustP, \ClustPP$ is independent of the structure of 
the remaining communities, because
any edge inside one of the other communities 
remains inside, and the \emph{expected number} of edges inside other
communities also stays the same. 
Thus, in splitting \Clust into \ClustP and \ClustPP, the 
modularity \MOD increases by
\begin{eqnarray*}
\ModChange{\Clust} & = & \frac{1}{m} \left(
\frac{(\sum_{v \in \ClustP} \Deg{v}) 
      (\sum_{u \in \ClustPP} \Deg{u})}{2m} 
- \SetCard{e(\ClustP,\ClustPP)}\right),
\end{eqnarray*}
where $e(\ClustP,\ClustPP)$ denotes the set of edges between \ClustP
and \ClustPP.

The target communities $\ClustP, \ClustPP$ are
calculated
using the above VP rounding, and the algorithm will terminate when none of the 
\ModChange{\Clust} are positive. The full algorithm is given below.

The use of a Max-Heap is not strictly necessary; a set 
of active communities would have been sufficient. However, the choice of a 
Max-Heap has the added advantage that by slightly tweaking the termination 
condition (requiring an increase greater than some $\epsilon$), one can force 
the communities to be larger, and the algorithm to terminate faster.

It is important that in each iteration of the algorithm, the degrees
$\Deg{v}$ for each vertex $v$ and the total number of edges $m$ be
calculated by taking into account all the edges in the entire graph and not
just the edges belonging to the sub-graph being partitioned.

\begin{algorithm}[H]
\caption{Hierarchical Clustering}
\begin{algorithmic}[1] 
	\STATE Let $M$ be an empty Max-Heap. 
	\STATE Let \Clust be a cluster containing all the vertices. 
	\STATE Use VP rounding to calculate (approximately) the maximum increase in 
	modularity possible, \ModChange{\Clust}, achievable by dividing \Clust into two 
	partitions.
	\STATE Add 
	$(\Clust,\ModChange{\Clust})$ to $M$.
	 \WHILE{the head element in $M$ has $\ModChange{\Clust} > 0$}
		\STATE Let \Clust be the head of $M$. 
	  	\STATE Use VP rounding to split \Clust into two partitions $\ClustP,
	 	\ClustPP$, and calculate $\ModChange{\ClustP}, \ModChange{\ClustPP}$.
		\STATE Remove \Clust from $M$. \STATE Add $(\ClustP, \ModChange{\ClustP}), 
		(\ClustPP, \ModChange{\ClustPP})$ to $M$.
	\ENDWHILE 
	\STATE Output as the final partitioning all the partitions remaining in the 
	heap $M$, as well as the hierarchy produced.
\end{algorithmic}
\end{algorithm}

As a post-processing step, we run the \emph{local-search} algorithm 
proposed by Newman \cite{newman:eigenvectors}. 
The post-processing brings the VP results nearly to par with those
obtained by the LP method. 


\subsection{Local Search Algorithm} \label{sec:post-processing}
We use the local-search algorithm proposed by Newman \cite{newman:eigenvectors} 
for refining the results obtained by our LP and VP methods.
This method improved the modularity of the partitions produced by
the LP method by less than 1\% and in the case of the QP method, it
improved the modularity by less than 5\%.
The local search method is based 
on the Kernighan-Lin algorithm for graph bisection
\cite{kernighan:lin}. Starting from some initial network clustering,
the modularity is iteratively improved as follows:
select the vertex which, when moved to another group, results in the
maximum increase in modularity (or minimum decrease, if no increase is
possible). 
In one complete iteration, each vertex changes its group exactly once;
at the end of the iteration, the intermediate clustering with the
highest modularity value is selected as the new clustering. 
This process is continued as long as there is an increase in the overall
modularity. For details of the implementation, we refer the reader to
\cite{newman:eigenvectors}.

%
%

\section{Examples} \label{sec:experiments}
In this section, we present results for both of our algorithms on several 
real-world networks. We focus on well-studied networks since our goal in this 
paper is to compare the quality of optimization achieved by our methods to 
approaches in past work, rather than discovering novel structure. We restrict 
our attention here to networks with at most a few thousand nodes, as this is 
currently the limit for our algorithms. The algorithm implementations 
are available online at \texttt{http://www-scf.usc.edu/\~{}gaurava}.



We evaluate our results in two ways: manually and by comparing against past 
work. For several smaller networks, we show below the clusterings obtained by 
the LP rounding algorithm. In all of the cases, the clusterings can be seen to 
be closely correlated with some known ``semantic'' information about the 
network. 

\subsection{Zachary's Karate Club}

\begin{figure} \includegraphics[width=5cm]{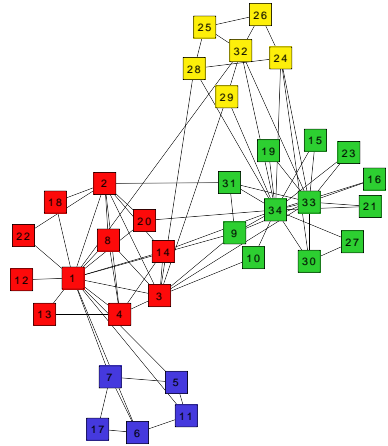} 
\caption{The optimal 
community structure with modularity $0.4197$ for Zachary's Karate Club 
network. Each community is shaded with a different color.}
\label{fig:karate}
\end{figure}

The \emph{``Zachary's Karate Club''} \cite{zachary:karate} network
represents the friendships between 
34 members of a karate club in the US over a period of 2 years. It has 
come to be a standard test network for clustering algorithms, partly due to the 
fact that during the observation period, the club broke up into two separate 
clubs over a conflict, and the resulting two new clubs can be considered a 
``ground truth'' clustering. Both of our algorithms find a community 
structure identical to the one detected by Medus et
al.~\cite{medus:acuna:dorso:global-optimization}. 
It has a modularity of $0.4197$. Our algorithm also 
proves this value to be best possible, because the LP returned a
 $\SET{0,1}$-solution, i.e., no rounding was necessary.
The community structure found for the Karate Club network is shown in 
Figure \ref{fig:karate}.


For finding the primary two-community division in this network, we ran a single 
iteration of the VP algorithm and found a partition identical to that found by 
Medus et al.~\cite{medus:acuna:dorso:global-optimization}. This partition 
corresponds almost exactly to the actual factions in the club, with the 
exception of node 10. The bipartition found by the VP method has a modularity 
of 0.3718, whereas the partition corresponding to the actual factions in the 
club has a lower modularity of 0.3715. This explains the ``misclassification'' 
of node 10, and also emphasizes that no clustering objective can be 
guaranteed to always recover the ``semantically correct'' community structure 
in a real network. The latter should be taken as a cautioning against accepting 
modularity-maximizing clusterings as ground truth.

\subsection{College Football}
\begin{figure}[h]
 \includegraphics[width=6cm]{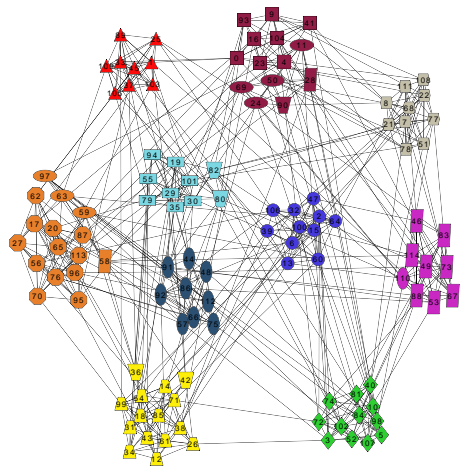} 
 \caption{The partitioning of the College Football 
 network found by the LP rounding algorithm. Each detected community is shaded 
 with a different color. The actual conferences are depicted using
 different shapes. \label{fig:foot_1}}
\end{figure}

This data set representing the schedule of Division I football games for the 
2000 season was compiled by Girvan and Newman \cite{girvan:newman}. 
Vertices in the graph represent teams, and edges represent regular 
season games between the two teams they connect. The teams are divided into 
\emph{conferences} with 8--12 teams each. Usually, more games are played within 
conferences than across conferences, and it is an interesting question whether 
the ground truth of conferences can be reconstructed by observing the games 
played. Both our algorithms find the
same clustering with modularity $0.6046$, shown in Figure \ref{fig:foot_1}.
The algorithms accurately recover most of the conferences as well as 
the independent teams (which do not belong to any conference).


Our algorithms also found a slightly suboptimal clustering of
modularity $0.6044$, combining two prominent conferences,
\emph{Mountain West} and \emph{Pacific 10} (brown squares and gray
hexagons in the top right corner) into one community. 
The reason is that many games were played between teams of 
the two conferences. This shows that community detection is inherently 
unstable: solutions with only slightly different modularity (differing only by 
$0.0002$) can differ significantly. Such slight differences could easily elude 
heuristic algorithms. More importantly, this instability shows again that 
communities maximizing modularity should be evaluated carefully for semantic 
relevance.
With respect to such instabilities in community structure, Gfeller et
al.~\cite{gfeller:chappelier:delosrios} give a more detailed analysis
and provide methods for detecting them. However, their methods are
applicable only to non-randomized clustering algorithms 

This example illustrates an advantage of our randomized rounding algorithms, 
which produce multiple different solutions. These solutions 
together often reveal more information about community
  boundaries. They can also be 
manually inspected if desired, and a researcher with domain knowledge can pick 
the one representing the true underlying structure most accurately.

\subsection{Books on American Politics}
\begin{figure}[h]
\includegraphics[width=6cm]{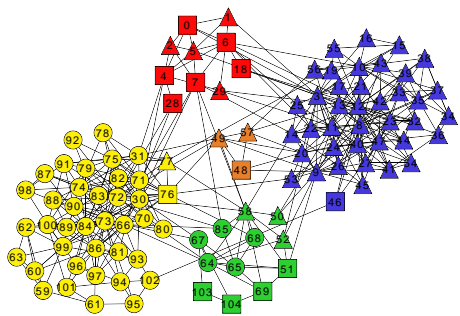}
\caption{The partitioning of the American Political
Books network found by the LP rounding algorithm. Each detected community is
shaded with a different color, while actual political slants are
depicted using different shapes. The circles are \emph{liberal books}, the
triangles are \emph{conservative books}, and the squares are \emph{centrist}.}
\label{fig:polbooks}
\end{figure}

As a final example, Figure \ref{fig:polbooks} shows the community structure 
detected in the \emph{American Political Books} network compiled by V.~Krebs. The 
vertices represent books on American politics bought from amazon.com, and edges 
connect pairs of books frequently co-purchased. The 
books in this network were classified by Newman 
\cite{newman:modularity-community} into categories liberal or conservative, 
except for a small number of books with no clear ideological leaning. Figure 
\ref{fig:polbooks} shows that our algorithm accurately detects a strong 
community  structure, which matches fairly well the underlying
semantic division based on political slant.

The community structure produced by our LP algorithm has a modularity
of $0.5272$ and agrees mostly with the manual labeling. It is very similar 
to the one produced by Newman \cite{newman:eigenvectors},
except for an extra cluster of three nodes produced by our method, as
well as slightly fewer ``misclassified'' nodes in the two main
clusters. The three books in the additional cluster were 
biographical in nature, and were always bought together.
The additional cluster is not found by the VP method, which instead
merges the three biographical books with the blue cluster, and obtains
a modularity of 0.5269.
 
For finding the primary division in this network, we ran a single iteration of 
the VP algorithm. The partition has a modularity value of 0.4569. 
It produces a partition with all the \emph{liberal books} and three of the
\emph{conservative books} assigned to one cluster and the remaining
\emph{conservative books} assigned to the other cluster. The \emph{centrist
books} were divided roughly evenly among the two clusters.

We also computed the modularity values for various ``ground truth'' 
partitionings. If the books are divided into three communities corresponding to 
liberal, conservative, and centrist (according to a manual labeling), the 
modularity is significantly inferior to our best clustering, namely 0.4149. 
If the centrist books are completely grouped with either the liberal or 
conservative books, the modularity deteriorates further to 0.3951
resp.~0.4088, which is noticeably worse than the modularity of 0.4569
achieved by the bipartition of our algorithm.
This corroborates an observation already made in discussing the Zachary Karate 
Club: the semantic ground truth partitioning will not necessarily 
achieve the highest modularity as a network partition, and hence, the two 
should not be treated as identical.

\subsection{Other Examples}
We tested our methods on several other networks and were able to identify 
community structures with very high modularity values. The test networks 
included a collaboration network of \emph{jazz musicians} (JAZZ) \cite{gleiser:danon},
the social network of a community of 62 \emph{bottlenose dolphins}
(DOLPH) living in Doubtful Sound, New Zealand \cite{lusseau:dolphin}, 
an interaction network of the characters from Victor Hugo's novel
\emph{Les Mis\'{e}rables} (MIS) \cite{knuth:graphbase}, 
a \emph{collaboration network} (COLL) of scientists who conduct
research on networks \cite{newman:collaboration-paths}, 
a \emph{metabolic} network for the nematode \emph{C.elegans}
(META) \cite{jeong:tomber:albert:oltvai:barabasi} and a network of \emph{email 
contacts} between students and faculty (EMAIL) 
\cite{guimera:danon:diaz-guilera:giralt:arenas}.

We compare our algorithms against past published partitioning 
heuristics, specifically, the edge-betweenness based 
algorithm of Girvan and Newman \cite{girvan:newman} (denoted by GN), the 
extremal optimization algorithm of Duch and Arenas \cite{duch:arenas} (DA) and 
the eigenvector based algorithm of Newman 
\cite{newman:eigenvectors,newman:modularity-community}. The bottom-up heuristic 
of Clauset, Moore, and Newman \cite{clauset:newman:moore} is designed not so 
much to yield close-to-optimal clusterings as to give reasonable clusterings 
for extremely large networks (several orders of magnitude beyond what our 
algorithms can deal with); the performance of their heuristic is significantly 
inferior to the other methods.

\begin{table}[h] \centering
\begin{small}
\begin{tabular}{|l | r | c | c | c | c | c | c|}
\hline Network & size 		$n$      	& GN    & DA  		& EIG   & VP    & LP    & UB\\
\hline KARATE   & 34       	& 0.401	& 0.419 	& 0.419 & 0.420	& 0.420 &0.420\\
\hline DOLPH   		& 62 		& 0.520	&    -		&   -   & 0.526 & 0.529 & 0.531\\
\hline MIS 			& 76 		& 0.540	&    -		&   -   & 0.560 & 0.560 & 0.561\\
\hline BOOKS			& 105 		&   -	&    -		& 0.526 & 0.527 & 0.527 & 0.528\\
\hline BALL  			& 115 		& 0.601 &    -		&   -   & 0.605 & 0.605 & 0.606\\
\hline JAZZ			& 198 		& 0.405	& 0.445 	& 0.442 & 0.445 & 0.445 & 0.446\\
\hline COLL    		& 235 		& 0.720	&    -		&   -   & 0.803 & 0.803 & 0.805\\
\hline META			& 453		& 0.403	& 0.434		& 0.435	& 0.450	&  -	&  -	\\
\hline EMAIL			& 1133		& 0.532	& 0.574		& 0.572	& 0.579	&  -	&  -	\\
\hline 
\end{tabular}
\caption{The modularity obtained by many of the previously published
methods and by the methods introduced in this paper, along with the upper
bound. 
\label{tab:results}}
\end{small}
\end{table}

Both the LP and VP rounding algorithms outperformed all other methods in terms 
of the value of modularity obtained. We summarize the results obtained by all 
algorithms as well as the upper bound (denoted by UB) in
Table~\ref{tab:results}. 
(Some LP heuristic and upper bound entries for larger data sets are
missing, because the LP solver could not solve such large instances.)
Notice that it is not clear whether the upper bound can in fact be
attained by any clustering. It is, however, striking how close to the
upper bound the clusterings found by the LP and VP rounding algorithms are.

\section{Conclusion}

We have shown that the technique of rounding solutions to fractional
mathematical programs yields high-quality modularity maximizing
communities, while also providing a
useful upper bound on the best possible modularity.

The drawback of our algorithms is their resource requirement. Due to
$\Theta(n^3)$ constraints in the LP, and $\Theta(n^2)$ variables in
the VP, the algorithms currently do not scale beyond about 300
resp.~4000 nodes. Thus, a central goal for future work would be to
improve the running time without sacrificing solution quality. An
ideal outcome would be a purely combinatorial algorithm avoiding the
explicit solution to the mathematical programs, but yielding the same
performance. 


Secondly, while our algorithms perform very well on all networks we considered, 
they do not come with \emph{a priori guarantee} on their performance. 
Heuristics with such performance guarantees are called
\todef{approximation algorithms}
\cite{vazirani:approximation-algorithms}, and are desirable because
they give the user a hard guarantee on the solution quality, even for
pathological networks. Since the algorithms 
of Charikar et al.~and Goemans and Williamson on which our 
approaches are based do have provable approximation guarantees,
one would hope that similar guarantees could be attained for
modularity maximization. However, this does not hold for the
particular algorithms we use, due to the shift of the objective
function by a constant. Obtaining approximation algorithms for
modularity maximization thus remains a challenging direction for
future work.

\subsubsection*{Acknowledgments}
We would like to thank Aaron Clauset, Cris Moore, Mark Newman and Ashish 
Vaswani for useful discussions and advice, and Fernando Ord\'{o}\~{n}ez for 
providing computational resources. We also thank several anonymous
reviewers for helpful feedback on a previous version of the paper.
David Kempe has been supported in part by NSF CAREER Award 0545855.

\bibliographystyle{plain}
\bibliography{names,conferences,publications,bibliography}
\end{document}